\documentclass[pra,twocolumn,showpacs]{revtex4}
\usepackage{amsmath}
\usepackage{graphicx}
\def\ket#1{\left|#1\right\rangle}
\def\bra#1{\left\langle#1\right|}

\begin{document}

\title{Dark-State Polaritons in Single- and Double-$\Lambda$ Media}

\author{Y.~D.~Chong}
\email{cyd@mit.edu}
\author{Marin Solja\v{c}i\'{c}}

\affiliation{Department of Physics, Massachusetts Institute of
  Technology, Cambridge, Massachusetts 02139}

\date{\today}

\begin{abstract}
We derive the properties of polaritons in single-$\Lambda$ and
double-$\Lambda$ media using a microscopic equation-of-motion
technique.  In each case, the polaritonic dispersion relation and
composition arise from a matrix eigenvalue problem for arbitrary
control field strengths.  We show that the double-$\Lambda$ medium can
be used to up- or down-convert single photons while preserving quantum
coherence.  The existence of a dark-state polariton protects this
single-photon four-wave mixing effect against incoherent decay of the
excited atomic states.  The efficiency of this conversion is limited
mainly by the sample size and the lifetime of the metastable state.
\end{abstract}


\maketitle

\section{Introduction}

Several years ago, Fleischhauer and Lukin \cite{Lukin} predicted the
existence of a stable polaritonic excitation in $\Lambda$-type media
[Fig.~\ref{lambda figure}(a)] exhibiting electromagnetically-induced
transparency (EIT) \cite{EIT}.  This excitation, which involves a
vanishing population of excited states and was therefore dubbed the
``dark-state polariton'' (DSP), is a coherent quantum excitation whose
evolution is governed by a classical control field.  This provides a
method for manipulating single-photon motion, including stopping light
\cite{Phillips,Hau}.  The Fleischhauer-Lukin result was derived as a
perturbation expansion of the field operator equations of motion in
the strong control field limit.  In a subsequent work, Juzeli\=unas
and Carmichael used a Bogoliubov-type transformation to diagonalize
the model Hamiltonian exactly, and showed that the DSP can be
understood as a part of a branch of slow polaritons occurring in
systems containing a pair of atomic resonances \cite{Juz}.  These
authors also emphasized the fact that the photonic part of the
polariton mixes with atomic excitations possessing wavevectors
differing by the wavevector of the control field.  Thus, for instance,
it is possible to reverse the direction of a polariton wavepacket by
switching the direction of the control field.

In this paper, we derive the properties of the DSP using the
Sawada-Brout technique \cite{SB}.  This can be thought of as a
simplified version of the method used by Juzeli\=unas and Carmichael,
and we shall see how its results reduce to those of Fleischhauer and
Lukin near resonance, which was not demonstrated in Ref.~\cite{Juz}.
We then extend the analysis to a double-$\Lambda$ medium
[Fig.~\ref{lambda figure}(b)], which contains a DSP consisting of
low-lying atomic excitations and photon states of two different
frequencies \cite{Raczynski,Li}.  In both single- and double-$\Lambda$
systems, the DSP is protected against incoherent decay processes
acting on the excited states, because it contains a vanishing
population of these states.  It has previously been shown that
double-$\Lambda$ media can efficiently upconvert classical probe beams
\cite{Korsunsky,Merriam}, and a related four-wave mixing scheme has
already been used in such systems to generate correlated photon pairs
\cite{Kolchin,Braje,Balic,Thompson}.  Here, we point out that the DSP
could be exploited to perform single-photon frequency conversion in a
manner that preserves quantum coherence.  Unlike semiclassical
analyses in which the electromagnetic field is treated classically
\cite{Korsunsky,Merriam}, this theory applies to the single-photon
regime.  It may thus have applications in quantum information
processing, such as for downconverting a member of an entangled photon
pair to a frequency suitable for transmission over a
telecommunications fiber.  Unlike parametric conversion schemes
exploiting optical nonlinearities, the relevant photons are up- or
down-converted individually, instead of being split or recombined; the
additional momentum and energy are supplied by the control fields.

\section{Single-$\Lambda$ System}

\begin{figure}
\centering
\includegraphics[width=0.35\textwidth]{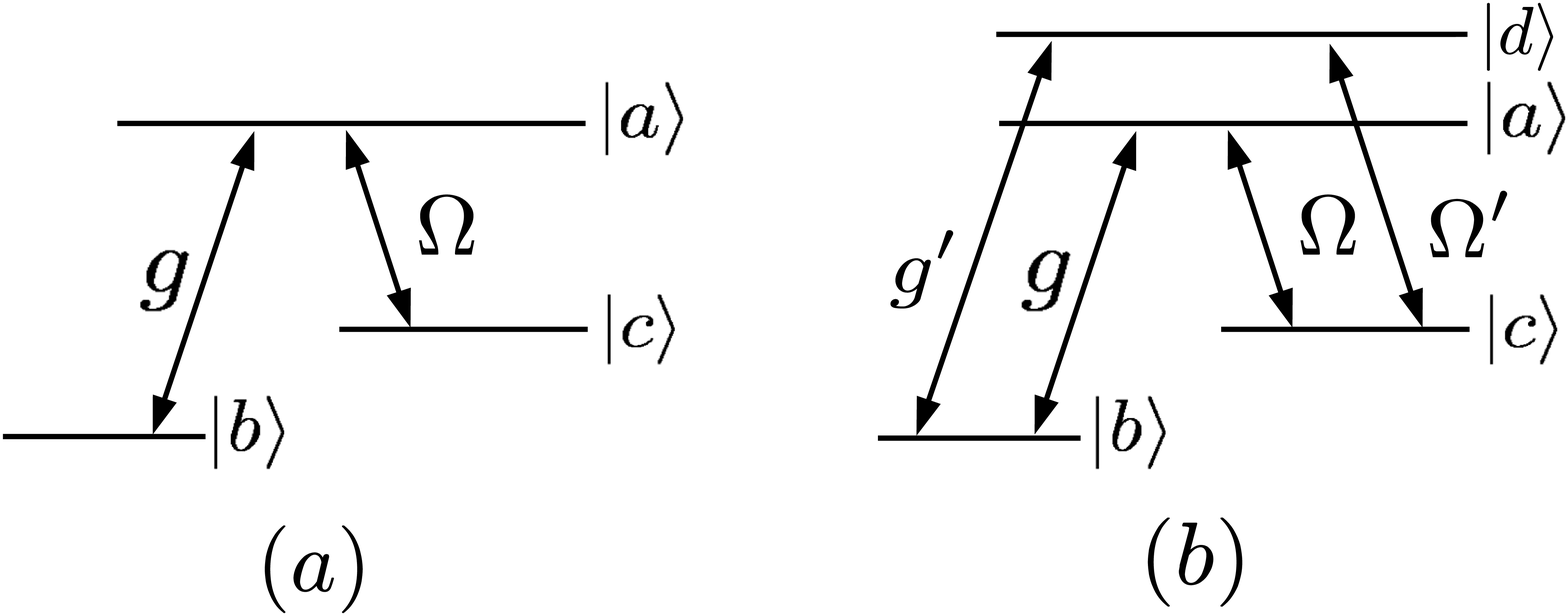}
\caption{(a) 3-level $\Lambda$-type medium.  (b) Double-$\Lambda$
  medium.}
\label{lambda figure}
\end{figure}

We begin by considering an $N$-atom gas with a single-$\Lambda$ level
structure, shown in Fig.~\ref{lambda figure}(a).  The ground, excited,
and metastable atomic states are respectively denoted by $\ket{b}$,
$\ket{a}$, and $\ket{c}$, and their corresponding energies by
$\hbar\omega_b$, $\hbar\omega_a$, and $\hbar\omega_c$.  The atomic
Hamiltonian is
\begin{equation}
  H_0 = \hbar \sum_r \left(\omega_a \sigma^{aa}_r + \omega_b
  \sigma^{bb}_r + \omega_c \sigma^{cc}_r \right),
\end{equation}
with the sum performed over all atomic positions $r$.  Here,
\begin{equation}
  \sigma^{\mu\nu}_r \equiv \ket{\mu}_r \bra{\nu}_r
\end{equation}
denotes a transition operator for the atom at position $r$.  We also
define Fourier-transformed operators, $\sigma^{ab}_k \equiv N^{-1/2}
\sum_r \sigma^{ab}_r\,e^{ikr}$ etc.  The photon Hamiltonian is
\begin{equation}
  H_1 = \sum_{k} \hbar c |k| a_k^\dagger a_k,
\end{equation}
where $a_k^\dagger$ and $a_k$ are photon creation and destruction
operators.  The photons interact with the $ab$ transition through the
minimal-coupling Hamiltonian
\begin{equation}
  H_2 = - \hbar g \; \sum_{k} \,a_k \,\sigma^{ab}_{k}
  +\,\textrm{h.c.}
  \label{atom-photon interaction}
\end{equation}
The coupling constant is $g \simeq \mathcal{P} \sqrt{2\pi N
  \omega_{ab} / \hbar V}$, where $\mathcal{P}$ is the dipole moment of
the $ab$ transition, $\omega_{ab}\equiv\omega_a-\omega_b$, and $V$ is
the cavity volume.  For notational simplicity, we have used the
rotating-wave approximation.  Finally, we include a classical control
field with strength (Rabi frequency) $\Omega$, frequency $\omega_L
\sim \omega_{ac}$, and wavevector $k_L$:
\begin{equation}
  H_3(t) = - \hbar \Omega \, e^{-i\omega_Lt} \, \sum_r  \, e^{ik_L r} \,
  \sigma^{ac}_r + \, \textrm{h.c.}
  \label{H3}
\end{equation}
Here, we have again discarded counter-rotating terms.  We neglect the
coupling between the photons and the $ac$ transition, which is
negligible compared to the effects of the control field, and the
coupling between the control field and the $ab$ transition, which is
off resonance.

The time dependence in (\ref{H3}) can be removed by defining
\begin{equation}
  H_L = U_L(t) H(t) U_L^\dagger(t) + \hbar \omega_L \sum_r \sigma^{cc}_r,
  \label{H_L}
\end{equation}
where $H(t) \equiv H_0 + \cdots + H_3(t)$, and
\begin{equation}
  U_L(t) = \exp\left[- i\omega_Lt \, \sum_r \sigma^{cc}_r\right].
  \label{U_L}
\end{equation}
The Schr\"odinger equation
$H(t)\ket{\psi(t)}=i\hbar\partial_t\ket{\psi}$ can be then rewritten
as
\begin{equation}
  i\hbar \frac{\partial}{\partial t} \,
  \left[U_L(t)\ket{\psi(t)}\right] = H_L
  \left[U_L(t)\ket{\psi(t)}\right].
  \label{modified Schrodinger equation}
\end{equation}
Thus, we can extract solutions to the Schr\"odinger equation from the
energy eigenstates of the time-independent Hamiltonian $H_L$.  To
obtain these, we look for a polariton excitation operator $A^\dagger$
such that
\begin{equation}
  [H_L, A^\dagger] = \hbar\omega A^\dagger + \cdots
  \label{commutation}
\end{equation}
The polariton is long-lived provided the omitted terms are negligible
\cite{SB}.  If the initial state of the system is its (zero photon)
ground state, $A^\dagger$ should be a mixture of $a^\dagger$,
$\sigma^{ab}$, and $\sigma^{cb}$.  Below, we list the commutation
relations of these three operators with $H_L$.  We have removed terms
involving $\sigma^{ba}$, $\sigma^{aa}$, $\sigma^{ca}$, and $a_k$;
since these operators give zero when acting on the ground state, this
introduces no additional error for single-polariton excitations.
Similarly, we have replaced $\sigma^{bb}_k$ with
$\sqrt{N}\delta_{k0}$.  Thus,
\begin{eqnarray}
\left[H_L, \sigma^{ab}_k\right] &\simeq& \hbar\omega_{ab} \sigma^{ab}_k -
\hbar\Omega^* \,\sigma^{cb}_{k-k_L} - \hbar g^* a_k^\dagger
\label{ab commutation}\\
\left[H_L, \sigma^{cb}_{k-k_L}\right] &\simeq&
\hbar(\omega_{cb}+\omega_L) \sigma^{cb}_{k-k_L} - \hbar\Omega
\,\sigma^{ab}_k
\label{cb commutation} \\
\left.[H_L, a_k^\dagger] \right. &=&
\hbar c |k| a_k^\dagger - \hbar g \,\sigma^{ab}_k.
\label{a commutation}
\end{eqnarray}

\begin{figure}
\centering
\includegraphics[width=0.36\textwidth]{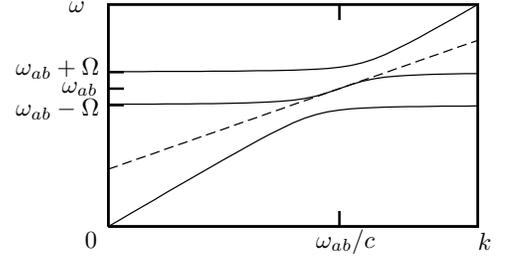}
\caption{Polaritonic dispersion curve for $\omega_L = \omega_{ac}$.
  The solid lines show the exact polariton solutions given by
  Eq.~(\ref{eig equation}); the dashed line showss the
  Fleischhauer-Lukin solution, Eq.~(\ref{polariton energy}).}
\label{dispersion curve}
\end{figure}

Let us now look for excitation operators of the form
\begin{equation}
  A_{nk}^\dagger = - \phi^1_{nk} \, \sigma^{ab}_k + \phi^2_{nk} \,
  \sigma^{cb}_{k-k_L} + \phi^3_{nk} \,a_k^\dagger,
  \label{polariton operator}
\end{equation}
where the band index $n$ enumerates the different polariton species.
The c-numbers $\phi^j_{nk}$ are determined by inserting
(\ref{polariton operator}) into (\ref{commutation}) and using (\ref{ab
  commutation})-(\ref{a commutation}).  This gives three
self-consistency equations that can be written as
\begin{equation}
\begin{bmatrix}
  \omega_{ab} & \Omega & g \\
  \Omega^* & \omega_{cb}+\omega_L & 0 \\
  g^* & 0 & c|k|
\end{bmatrix}
\begin{bmatrix}
\phi^1 \\ \phi^2 \\ \phi^3
\end{bmatrix}_{nk}
= \omega_{nk}
\begin{bmatrix}
\phi^1 \\ \phi^2 \\ \phi^3
\end{bmatrix}_{nk}.
\label{eig equation}
\end{equation}
The form of the effective Hamiltonian in (\ref{eig equation}) is
familiar from semiclassical analyses of EIT.  Fig.~\ref{dispersion
  curve} shows the bandstructure in the absence of loss, similar to
the one given in Ref.~\cite{Juz}.  For simplicity, let us assume that
$\omega_L = \omega_{ac}$.  The asymptotic eigenfrequencies far from
resonance are $c |k|$ and the eigenvalues of the upper-left $2\times2$
submatrix in the effective Hamiltonian, in this case $\omega_{ab} \pm
\Omega$.  Exactly at resonance ($|k| = \omega_{ab}/c$), there is an
eigenvector $\propto [0,1,-\Omega/g]$, and for slightly detuned $k$
this eigenvector continues into ones where the $\sigma^{ab}$ component
is nonzero but small.  These solutions---``dark-state
polaritons''---are thus insensitive to incoherent decay processes
acting on $\ket{a}$.  The stability of the exactly-resonant DSP is
limited only by the lifetime of the metastable state $\ket{c}$, which
we shall assume to be longer than the time-scale of any relevant
experiment.  For off-resonant DSPs, the decay rate is only quadratic
in the detuning: upon replacing $\omega_a$ with $\omega_a - i
\Gamma_a$ in (\ref{eig equation}), one finds that the imaginary part
acquired by $\omega_{k}$ is $\sim \Gamma_a|\Delta/\Omega|^2$ (for
$\Omega \gg g$), where $\Delta \equiv c|k| - \omega_{ab}$.  The other
two polariton branches correspond to ``bright'' polaritons that
contain significant $\ket{a}$ population and are thus strongly
affected by losses.  As in Ref.~\cite{Lukin}, we neglect Langevin
noise effects, which do not influence the adiabatic evolution of the
DSPs.

Expanding around $\omega = \omega_{ab}$ yields a limiting solution for
the DSPs:
\begin{eqnarray}
  \omega_k &=& \omega_{ab} + \frac{|\Omega|^2}{|g|^2 + |\Omega|^2} \,
  (c|k| - \omega_{ab})
  \label{polariton energy}\\
  \phi^1_{k} &=& \frac{\Omega (c|k| - \omega_{ab})}{|g|^2 +
    |\Omega|^2} \, \phi^2_{k} \label{ab amplitude}\\
  \phi_{k}^3 &=& - \frac{\Omega}{g} \, \phi^2_{k}.
  \label{cb amplitude}
\end{eqnarray}
Eq.~(\ref{polariton energy}-\ref{cb amplitude}) agree with the
solution derived by Fleischhauer and Lukin using a perturbation
expansion in $1/\Omega$ \cite{Lukin}.  In our formalism, the fact that
decreasing $\Omega$ reduces the polaritonic group velocity can be
intuitively understood as the result of ``squeezing'' the bandwidth of
the middle polariton band.  An interesting property of the DSP
solution is that it does not depend on the energies of the underlying
$\Lambda$ system, only the detuning of the control field and the
coupling parameters $g$ and $\Omega$.

Finally, we can extract the solutions to the original Schr\"odinger
equation using (\ref{modified Schrodinger equation}).  For a polariton
with quantum numbers $(n,k)$, the state at time $t$ is
\begin{multline}
  \ket{\psi(t)} = e^{-i\omega_{nk}t} \\ \times \left[ - \phi^1_{nk} \,
    \sigma^{ab}_k + \phi^2_{nk} \, e^{i\omega_L t} \,
    \sigma^{cb}_{k-k_L} + \phi^3_{nk} \,a_k^\dagger \right] \ket{0}.
  \label{polariton state}
\end{multline}
The $\sigma^{cb}$ component in (\ref{polariton state}) has a different
frequency and wavevector compared to the rest of the polariton.  This
property does not, however, destabilize the polariton: in a wavepacket
constructed of a superposition of DSPs, the photonic and $\sigma^{cb}$
components possess different phase factors but share a single
envelope.

The preceding derivation holds regardless of the angle between the
input photon and the control beam.  The direction of $k_L$ only enters
into the choice of excitation operator $\sigma_{k-k_L}^{cb}$ occurring
in the polariton operator (\ref{polariton operator}), and plays no
role in the eigenproblem (\ref{eig equation}) that yields the state
amplitudes and polariton energy.

By switching between two non-collinear control beams, it is possible
to coherently rotate the photon wavevector, by an angle of up to
$2\sin^{-1}(\omega_{ac}/\omega_{ab})$, where the plane of rotation is
specified by the polarization of the control field.  A special case of
this has been discussed by Juzeli\=unas and Carmichael: when $\omega_b
\approx \omega_c$, one can coherently backscatter the photon by
inserting a photon with $k \sim k_L$, which mixes with a $\sigma^{bc}$
excitation with wavevector $k-k_L \sim 0$, and switching the control
field to $-k_L$.  The $\sigma^{bc}$ excitation then mixes into a
photon of wavevector $k - 2 k_L \sim -k$ \cite{Juz}.

\section{Double-$\Lambda$ System}

Suppose we add a second excited state, $\ket{d}$, as shown in
Fig.~\ref{lambda figure}(b).  A second control beam couples $\ket{d}$
to $\ket{c}$, and for simplicity we assume that the two control beams
have parallel polarization vectors.  The $d\leftrightarrow a$
transition is assumed to be forbidden.  One of the reasons this
``double-$\Lambda$'' system is interesting is that it can be used to
upconvert or downconvert probe beams, as experimentally demonstrated
by Merriam \textit{et.~al.}~\cite{Merriam} and other groups.  It can
be shown, using the Fleischhauer-Lukin formalism, that this type of
level structure supports a DSP \cite{Li}.  As we shall see, this DSP
arises naturally from the present method as a $5\!\times\!5$
generalizion of (\ref{eig equation}).

The Hamiltonian, $H'(t)$, contains four new terms.  The first, $\sum_r
\omega_d \sigma^{dd}_r$, gives the energy of the $\ket{d}$ states.
Next, we introduce a second photon field with operators $b_k^\dagger$
and $b_k$, and Hamiltonian $\sum_{k} \hbar c |k| b_k^\dagger b_k$.
(There is really only one photon field, but this trick is permissible
since the atom-photon coupling becomes negligible far away from the
EIT resonances.)  Finally, we add interaction terms analogous to
(\ref{atom-photon interaction}) and (\ref{H3}), with $\ket{d}$, $b_k$,
$g'$, and $\Omega'$ replacing $\ket{a}$, $a_k$, $g$, and $\Omega$
respectively.

\begin{figure}
\centering
\includegraphics[width=0.32\textwidth]{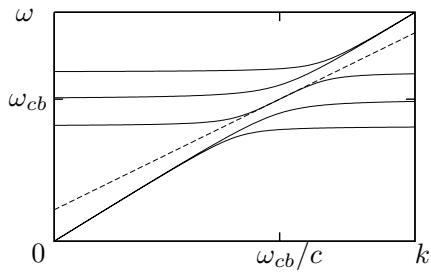}
\caption{Polaritonic dispersion curve for the double-$\Lambda$ medium.
  The solid lines show the exact polariton solutions given by
  Eq.~(\ref{new Hamiltonian matrix}); the dashed line shows the
  linearized solution given by Eq.~(\ref{omegank}).  The horizontal
  asymptotes occur at $\omega_{cb}$ and $\omega_{cb} \pm
  \sqrt{|\Omega|^2+|\Omega'|^2}$.}
\label{dispersion curve2}
\end{figure}

The control field interaction Hamiltonian (\ref{H3}) and its analog
for the $dc$ transition oscillate at different frequencies, so the
transformation (\ref{H_L})-(\ref{U_L}), which works by rotating
$\ket{c}$, cannot eliminate the time dependence.  We can overcome this
difficulty with a transformation that instead rotates the $\ket{a}$,
$\ket{d}$, and photonic states.  Let
\begin{widetext}
\begin{equation}
  H_L' = U_L'(t) H'(t) {U_L'}^\dagger - \hbar \omega_L \left({\sum}_k
  a^\dagger_k a_k + {\sum}_r \sigma^{aa}_r \right) - \hbar \omega_L'
  \left({\sum}_k b^\dagger_k b_k + {\sum}_r \sigma^{dd}_r \right),
\end{equation}
where $H'(t)$ is our new Hamiltonian, and
\begin{equation}
  U_L' = \exp\left[i\omega_Lt \left({\sum}_k a^\dagger_k a_k + {\sum}_r
    \sigma^{aa}_r\right) \right. \\
    \left. + i\omega_L't \left({\sum}_k b^\dagger_k b_k + {\sum}_r
    \sigma^{dd}_r\right) \right].
\end{equation}
This once again allows us to write the Schr\"odinger equation as
$i\hbar \partial_t \left[U_L'(t)\ket{\psi(t)}\right] = H_L'
\left[U_L'(t)\ket{\psi(t)}\right]$, where $H_L'$ is time-independent.
We look for excitation operators for $H_L'$ of the form
\begin{equation}
  A^\dagger_{nk} = - \phi^1_{nk} \sigma^{ab}_{k+k_L} - \phi^2_{nk}
  \sigma^{db}_{k+k_L'} + \phi^3_{nk} \sigma^{cb}_k + \phi^4_{nk}
  a^\dagger_{k+k_L} + \phi^5_{nk} b^\dagger_{k+k_L'}.
  \label{double lambda polariton}
\end{equation}
The self-consistency equations for the parameters $\phi^j_{nk}$ take
the same matrix form as (\ref{eig equation}), with effective
Hamiltonian
\begin{equation}
  \mathcal{H}_{k}' = \hbar \,
\begin{bmatrix}
  \omega_{ab} - \omega_L & 0 & \Omega & g & 0 \\
  0 & \omega_{db} - \omega_L' & \Omega' & 0 & g' \\
  \Omega^* & {\Omega'}^* & \omega_{cb} & 0 & 0 \\
  g^* & 0 & 0 & c|k+k_L| - \omega_L & 0 \\
  0 & {g'}^* & 0 & 0 & c|k+k_L'| - \omega_L' \\
\end{bmatrix}.
\label{new Hamiltonian matrix}
\end{equation}
\end{widetext}
The polariton created by (\ref{double lambda polariton}) is a valid
excitation because, as in the single-$\Lambda$ case, no extra
non-negligible terms are generated by commutating this operator with
the Hamiltonian.  Let us now assume that the control fields are
resonant, i.e.~$\omega_L = \omega_{ac}$ and $\omega_L' = \omega_{dc}$.
For $|k+k_L| = \omega_{ab}/c$ and $|k+k_L'| = \omega_{db}/c$, the
effective Hamiltonian (\ref{new Hamiltonian matrix}) has an
eigenvector $\propto [0,0,1,-\Omega/g,-\Omega'/g']$.  The first two
components of this eigenvector, corresponding to the two excited
states, are identically zero, so this represents a DSP consisting of
$\sigma^{cb}_k$ excitations and photons with wavevectors $k+k_L$ and
$k+k_L'$.  It can be shown that no other linearly independent
eigenvector with this property exists, so there is only one such DSP
solution.  The linearized DSP solution, analogous to (\ref{polariton
  energy})-(\ref{cb amplitude}), is
\begin{eqnarray}
  \omega_{nk} &=& \omega_{cb} + \frac{|\Omega/g|^2 \delta k
    +|\Omega'/g'|^2 \delta k'}{1+|\Omega/g|^2+|\Omega'/g'|^2}
   \label{omegank}\\
  \phi^1_{nk} &=& \frac{\Omega}{|g|^2} \, \frac{\delta k +
    |\Omega'/g'|^2(\delta k - \delta
    k')}{1+|\Omega/g|^2+|\Omega'/g'|^2} \, \phi^3_{nk}
  \label{phi1}  \\
  \phi^2_{nk} &=& \frac{\Omega'}{|g'|^2} \, \frac{\delta k' +
    |\Omega/g|^2(\delta k' - \delta k)}{1+|\Omega/g|^2+|\Omega'/g'|^2}
  \, \phi^3_{nk} \label{phi2}\\
  \phi^4_{nk} &=& -(\Omega/g) \, \phi^3_{nk} \label{phi4}
  \\ \phi^5_{nk} &=& -(\Omega'/g')\, \phi^3_{nk}, \label{phi5}
\end{eqnarray}
where $\delta k \equiv |k+k_L| - \omega_{ab}/c$ and $\delta k'=
|k+k_L'| - \omega_{db}/c$.  The polaritonic bandstructure, in the
absence of loss, is shown in Fig.~\ref{dispersion curve2}.

These results can be shown to be consistent with the single-photon
limit of a semiclassical analysis of the double-$\Lambda$ medium given
by Korsunsky and Kosachiov \cite{Korsunsky}.  In this one-dimensional
model, where the electromagnetic field is treated classically, the
Heisenberg equations of motion for the atomic system possesses a
stationary ``dark state'' solution that is decoupled from the
electromagnetic field and is stable against spontaneous emission.
This dark state exists only if the background field (consisting of two
probe beams and two control beams) obeys certain frequency, amplitude,
and phase matching conditions.  The frequency-matching condition is
\begin{eqnarray}
  \omega - \omega_L = \omega' - \omega_L' &=& \omega_{cb},
  \label{korsunsky1}
\end{eqnarray}
where $\omega$ and $\omega'$ are the respective frequencies of the
probe beams resonant with the $ab$ and $db$ transitions.  This
equation is exactly satisfied by (\ref{double lambda polariton}), for
which $\omega = c (k+k_L)$ and $\omega' = c (k+k_L')$.  The physical
meaning of (\ref{korsunsky1}) is particularly easy to deduce in the
present theory: in the single-photon limit, the stationary state
corresponds to a polaritonic solution of the form (\ref{double lambda
  polariton}), for which the photonic components cannot take on
arbitrary frequencies because they are coherently mixed via the atomic
excitation $\sigma^{cb}_k$.  The amplitude-matching condition for the
semiclassical dark state is
\begin{equation}
  \frac{\mathcal{P} E}{\Omega} = \frac{\mathcal{P}' E'}{\Omega'},
  \label{korsunsky2}
\end{equation}
where $\mathcal{P}$ and $\mathcal{P}'$ are the dipole moments for the
$ab$ and $db$ transitions, and $E$ and $E'$ are the electric field
amplitudes of the associated probe beams.  The electric field
amplitudes can be related to the quantum mechanical photon amplitudes
$\phi^4$ and $\phi^5$ by
\begin{eqnarray}
  E &\leftrightarrow& \sqrt{2\pi\hbar\omega_{ab}/V}\, \phi^4 \\
  E' &\leftrightarrow& \sqrt{2\pi\hbar\omega_{db}/V} \,\phi^5,
\end{eqnarray}
which can be verified by computing the expectation value
$\langle|E|^2\rangle$ produced by each photon creation operator.  With
this identification, the linearized DSP amplitudes (\ref{phi4}) and
(\ref{phi5}) satisfy (\ref{korsunsky2}).  The third condition derived
by Korsunsky and Kosachiov, which relates the phases of the four
beams, is also satisfied by the DSP because, as shown by (\ref{phi4})
and (\ref{phi5}), the phases of the probe beams are locked to those of
the control beams $\Omega$ and $\Omega'$.

The dark state studied by Korsunsky and Kosachiov is a pure state of
the atomic system, reflecting the fact that the electromagnetic field
is treated classically \cite{Korsunsky}.  In contrast, the present
model takes into account the coherent mixing between the quantum state
of the probe field and the quantum state of the atomic medium:
performing a partial trace of the DSP over the photonic Hilbert space
yields a mixed atomic state.  This mixing becomes important at the
single-photon level, which is also potentially the regime of interest
for quantum information processing.  In the following section, we will
examine how this mixing can be used to convert between the two
photonic components of the double-$\Lambda$ DSP.

\section{Frequency Conversion}

For a single-$\Lambda$ medium with a resonant control beam, inserting
a photon with wavevector $k_0$, resonant with the $ab$ transition,
gives rise to a DSP whose group velocity points in the same direction,
independent of the direction of the control beam.  This freedom to
choose the direction of the control beam disappears in the
double-$\Lambda$ case.  Here, an incident photon $k_0$ mixes with
another photon with wavevector $k_1 = k_0 - k_L + k_L'$.  Assuming
both control beams are tuned to resonance, the resulting state
overlaps with a DSP only if $|k_1| \simeq \omega_{db}/c$.
Furthermore, the group velocity of the DSP is, from (\ref{omegank}),
\begin{equation}
  v = \nabla_k \omega_{nk} = \frac{|\Omega/g|^2 \; \hat{k}_0 +
    |\Omega'/g'|^2 \; \hat{k}_1} {1+|\Omega/g|^2+|\Omega'/g'|^2},
  \label{group velocity}
\end{equation}
where $\hat{k}_0 = k_0/|k_0|$ and $\hat{k}_1 = k_1/|k_1|$.  Therefore,
a choice of $\hat{k}_0$ and $\hat{k}_1$ determines the directions of
the two control beams (or, more generally, choosing any two of these
directions determines the other two).  As an aside, we note that the
beam matching conditions forbid the choice $\hat{k}_0 = -\hat{k}_1$,
which would imply the possibility of a stationary wavepacket with
nonzero control beams; however, if $\hat{k}_0$ and $\hat{k}_1$ are
\textit{nearly} antiparallel, (\ref{group velocity}) predicts that the
control beam strengths can be tuned to produce a low group velocity.

\begin{figure}
\centering
\includegraphics[angle=270,width=0.458\textwidth]{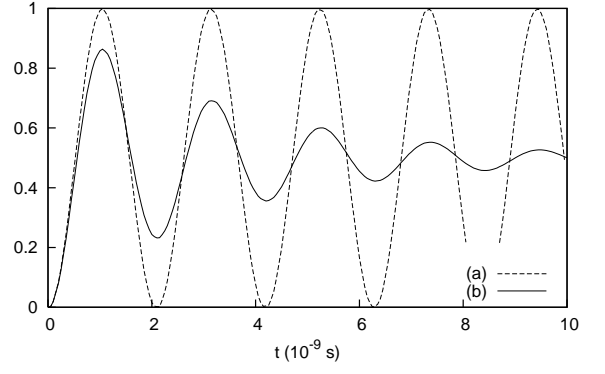}
\caption{Numerical solutions of $|\bra{0}b_{k-k_L+k_L'}\ket{\psi(t)}|$
  against $t$, where $\ket{\psi(t)}$ is the quantum state at time $t$
  after inserting a photon $a_k^\dagger$ with $k = \omega_{ab}/c$.
  Here, $\omega_{cb} = 10^4\,\textrm{cm}^{-1}$, $|g| = |g'| =
  0.1\,\textrm{cm}^{-1}$, and $|\Omega| = |\Omega'| =
  1\,\textrm{cm}^{-1}$.  (a) No excited state decay, $\Gamma_a =
  \Gamma_d = 0$.  (b) $\Gamma_a = \Gamma_d =
  0.02\,\textrm{cm}^{-1}$. }
\label{upconversion figure}
\end{figure}

In order to illustrate the mixing between the two photonic components
in the DSP, let us fall back on the ``trivial'' one-dimensional case
where all wavevectors are parallel, which satisfies the above beam
matching conditions.  Suppose we inject the photon $k_0$ at $t=0$, so
that the quantum state is
\begin{equation}
  \ket{\psi(0)} = a^\dagger_{k_0} \ket{0} = \sum_{n=1}^5
  \phi^{*\,4}_{nk} A^\dagger_{nk} \ket{0},
  \label{double lambda polariton operator}
\end{equation}
where $k\equiv k_0-k_L$.  Without losses, the state at time $t$ is
\begin{equation}
  \ket{\psi(t)} = e^{i\omega_L' t} \left(e^{-i\mathcal{H}_{k}' t /
    \hbar}\right)_{5,4} \; b^\dagger_{k+k_L'} \ket{0} + \cdots,
\end{equation}
where the matrix $\mathcal{H}$ is defined in (\ref{new Hamiltonian
  matrix}) and the omitted terms are the other polariton components.
The result, shown in Fig.~\ref{upconversion figure}, is an oscillating
upconversion amplitude $|\bra{0}b_{k+k_L'}\ket{\psi(t)}|$ that can
approach 100\%.  The effects of incoherent excited state decay, which
can be modeled by replacing $\omega_a$ with $\omega_a - i \Gamma_a$
and $\omega_d$ with $\omega_a - i \Gamma_d$ in (\ref{new Hamiltonian
  matrix}), are also shown in Fig.~\ref{upconversion figure}.
Although the DSPs are protected against decay, damping still occurs
because the incident photon generates a non-vanishing population of
bright polaritons.  When these exit the system (typically as off-axis
photons), only the DSP remains.

\begin{figure}
\centering
\includegraphics[width=0.46\textwidth]{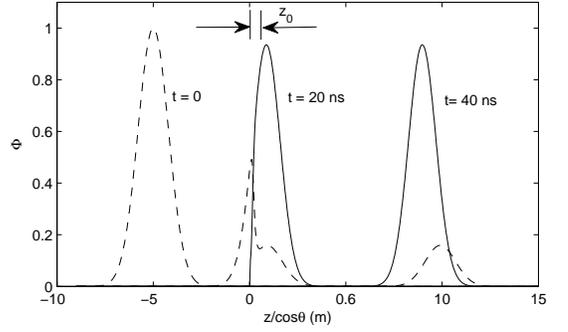}
\caption{Photon frequency conversion.  The $a$-photon amplitude
  $|\phi_4(z,t)|$ (dashed line) and $b$-photon amplitude
  $|\phi_5(z,t)|$ (solid line) are plotted at three instants.  The
  abscissa is $z/\cos\theta$, where $k_L \cdot \hat{z} = \cos\theta$.
  The second control beam is pointed that $k_L' \cdot \hat{z} = 0.9
  \cos\theta$.  Both control beams are c.w.  The effective thickness
  of the double-$\Lambda$ medium, $z_0 = 60\,\textrm{cm}$, is
  indicated.  Within the medium, $|g| = |g'| = 0.1\,\textrm{cm}^{-1}$,
  $\Gamma_a = \Gamma_d = 0.2 g$, $|\Omega| = 1\,\textrm{cm}^{-1}$, and
  $|\Omega'| = 3\{1+\tanh[4 (z/\cos\theta - 0.5) /
    z_0]\}\,\textrm{cm}^{-1}$.  Outside the sample lies vacuum.  The
  amplitudes are computed by integrating Eq.~(\ref{wave equation})
  numerically for the initial conditions in Eq.~(\ref{initial}), where
  $\beta = 0.8\,\textrm{m}^{-2}$ and $z_2 = - 50\,\textrm{cm}$.}
\label{space figure}
\end{figure}

\begin{figure}
\centering
\includegraphics[width=0.4\textwidth]{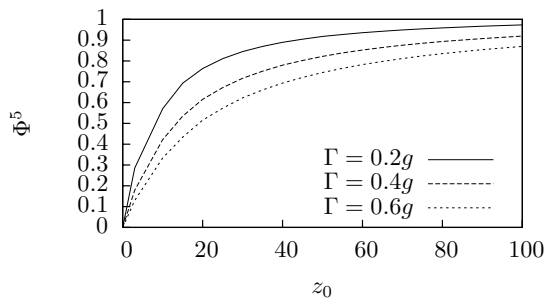}
\caption{Frequency conversion efficiency, parameterized by the
  converted photon amplitude $\Phi^5$ normalized to the input photon
  amplitude, as a function of the sample length $z_0$.  Within the
  sample, the control beam $\Omega'$ varies as $|\Omega'| =
  3\{1+\tanh[4 (z/\cos\theta - 0.5) / z_0]\}\,\textrm{cm}^{-1}$.
  Curves for incoherent decay rates $\Gamma_a = \Gamma_d = 0.2 g$,
  $0.4g$, and $0.6g$ are shown.  All other parameters are the same as
  in Fig.~\ref{space figure}.}
\label{efficiency figure}
\end{figure}

A more efficient example of single-photon frequency conversion can be
obtained by going from momentum space to real space and studying the
behavior of polariton wavepackets.  Let us define c-number fields
$\Phi_j = \Phi_j(r,t)$ such that
\begin{multline}
  \ket{\psi(t)} = \sum_r e^{i(\kappa r - \omega_{cb}t)} \left( -
  \Phi_1\, e^{ik_Lr} \sigma_{r}^{ab} - \Phi_2 \, e^{ik_L'r}
  \sigma_{r}^{db} \right.\\ \left. + \Phi_3 \, \sigma_{r}^{cb} +
  \Phi_4 \, e^{-ik_Lr} a^\dagger_r + \Phi_5 \, e^{-ik_L'r} b^\dagger_r
  \right).
  \label{real space state}
\end{multline}
Inserting (\ref{real space state}) into the Schr\"odinger equation and
using (\ref{new Hamiltonian matrix}), we obtain a Schr\"odinger
\textit{wave} equation
\begin{equation}
  i \hbar \frac{\partial\Phi_i}{\partial t} (r,t) = \sum_{j}
  \mathcal{H}_{ij} \; \Phi_j(r,t).
  \label{wave equation}
\end{equation}
If $\kappa$ is chosen such that $|\kappa+k_L| = \omega_{ab}/c$ and
$|\kappa+k_L'| = \omega_{db}/c$, then the DSP corresponds to values of
$\Phi_j$ that are constant in space.  For a wavepacket centered around
$\kappa$ with bandwidth $\ll \omega_{ab}, \, \omega_{db}$ (i.e.,
spatial width much longer than the optical wavelength, which is the
usual slowly-varying envelope approximation), $\mathcal{H}$ takes on
the intuitive local form
\begin{equation}
  \mathcal{H}(r,t) \approx \hbar \,
\begin{bmatrix}
  0 & 0 & \Omega & g & 0 \\
  0 & 0 & \Omega' & 0 & g' \\
  \Omega^* & {\Omega'}^* & 0 & 0 & 0 \\
  g^* & 0 & 0 & -ic\hat{k}_0\cdot\nabla & 0 \\
  0 & {g'}^* & 0 & 0 & -ic\hat{k}_1\cdot\nabla \\
\end{bmatrix}.
\label{wave Hamiltonian}
\end{equation}
As in the single-$\Lambda$ case, the evolution of the polaritonic
envelope is independent of the underlying double-$\Lambda$
frequencies, except through the coupling parameters $g$, $g'$,
$\Omega$, and $\Omega'$.  We again emphasize that this result is not
perturbative; it holds for arbitrary values of $\Omega$ and $\Omega'$,
and depends only on the fact that the wavepacket is sufficiently
broad.  Generally, the coupling parameters can vary (smoothly) in
space; for instance, a variation in $\Omega$ or $\Omega'$ could be
accomplished using a c.w.~control beam with a non-uniform
cross-sectional intensity profile.  Such variations can be used to
``adiabatically'' transfer one photon population to another within a
propagating DSP wavepacket, substantially improving the efficiency of
the conversion process compared to the previous example.

Let us consider an effectively one-dimensional experimental setup
where all relevant spatial variations occur in the $z$ direction.  In
particular, we must assume that the $x$ and $y$ edges are far enough
away that boundary effects (which appear when the beams are not all
collinear) are negligible.  The incident envelope field $\Phi_j(z,
t=0)$ is
\begin{align}
\begin{aligned}
\Phi_4 &=
\exp\left[-\beta\left(\frac{z}{\cos\theta} - z_2\right)^2\right], \\
\Phi_1 &= \Phi_2 = \Phi_3 = \Phi_5 = 0.
\label{initial}
\end{aligned}
\end{align}
Outside the sample ($z<0$ or $z>z_0$), all coupling parameters are
zero.  Within the sample ($0 < z < z_0$), the functional forms of
$\Omega(z)$ and $\Omega'(z)$ are chosen so that $|\Omega|\gtrsim |g|
\gtrsim |\Omega'|$ near the entrance of the sample, which ensures that
the DSP is dominated by the input photon; whereas $|\Omega'| \gtrsim
|\Omega|\gtrsim |g|$ near the exit, which ensures that the DSP is
dominated by the converted photon.  The result is shown in
Fig.~\ref{space figure}.  For the given parameters, the converted
photon amplitude is $\sim 0.9$ times the incident amplitude.  The
efficiency is limited by the available length of the double-$\Lambda$
medium.  As shown in Fig.~\ref{efficiency figure}, a longer sample
allows the $\Omega'$ field to be varied more gently, generating fewer
bright state polaritons and increasing the conversion efficiency.

\section{Conclusions}

In this paper, we have presented an analysis of single- and
double-$\Lambda$ EIT systems based on a microscopic equation-of-motion
technique.  Within this formalism, the presence of a DSP corresponds
to the existence of special eigenvectors of an effective Hamiltonian
matrix, in which the entries corresponding to rapidly decaying
excitations are identically zero, regardless of the strength of the
control fields.  The ability of the double-$\Lambda$ system to
efficiently upconvert and downconvert photons, previously established
in semiclassical four-wave mixing studies \cite{Korsunsky,Merriam}, is
retained in the coherent single-photon limit due to the existence of
the DSP.  The analysis can be further generalized to multi-$\Lambda$
systems, where one finds additional polaritonic bands similar to those
in Fig.~\ref{dispersion curve2}, with exactly one family of DSP
solutions possessing vanishing excited state populations.

Throughout this paper, we have restricted our attentions to the
single-polariton sector of the theory, which is valid only if the
polaritons are much more dilute than the underlying atomic medium.
The polariton operators (\ref{polariton operator}) and (\ref{double
  lambda polariton operator}) do not obey exact bosonic commutation
relations, since the $\sigma$ operators are not bosonic operators;
however, the corrections to the commutator vanish as $O(M/N)$, where
$M$ is the number of atoms excited \cite{Hopfield}.  This condition is
satisfied, for instance, in the experiments of Merriam
\textit{et.~al.}, where $M/N \sim 10^{-3}$ \cite{Merriam}.  The
single-polariton sector has the advantage that the quantum state of
the system can be expressed in terms of a simple wave equation, as in
(\ref{wave equation}).  Thus, once the $\sigma$ operators have been
used to derive the effective Hamiltonian, the additional structure
given by their non-trivial commuation relations disappears from the
theory.  Should one wish to study the limit where $M$ becomes
comparable to $N$, this structure will have to be taken into account.

We would like to thank P.~Bermel, S.~E.~Harris, J.~Taylor, and
V.~Vuletic for helpful comments.  This work was supported in part by
the Army Research Office through the Institute for Soldier
Nanotechnologies under Contract No.~W911NG-07-D-0004, and by DOE Grant
No.~DE-FG02-99ER45778.

\end{document}